\documentclass[showpacs,prb,preprintnumbers,twocolumn,amsmath,amssymb]{revtex4-2}
\usepackage{graphicx}% Include figure files
\usepackage{dcolumn}% Align table columns on decimal point
\usepackage{bm}% bold math
\usepackage{multirow}

\def\Cij{{C_{ij}}}
\def\Cone{{C_{11}}}
\def\Ctwo{{C_{12}}}
\def\C44{{C_{44}}}
\def\Cprime{{C^{\prime}}}
\def\T2g{{T_{2g}}}
\def\Eg{{E_{g}}}
\def\dxy{{d_{xy}}}
\def\dxz{{d_{xz}}}
\def\dyz{{d_{yz}}}
\def\EF{{E_F}}
\def\eg{{\em e.g. }}

\begin{document}

\title{Electronic structure and elasticity of the Ta-W solid solution}

\author{Kareem Abdelmaqsoud and John R. Kitchin}
\affiliation{Carnegie Mellon University, Department of Chemical Engineering, Pittsburgh, 15213, PA, USA}

\author{Michael Widom}
\affiliation{Carnegie Mellon University, Department of Physics, Pittsburgh, 15213, PA, USA}

\date{\today }

\begin{abstract}
The brittleness or ductility of metals has long been attributed to their elastic constants, with high Poisson ratio, or equivalently high Pugh ratio, favoring greater ductility. Growing evidence links ductility with their electronic structure. Consequently, it is desirable to understand how the electronic structure affects the elastic constants. Here, we examine the Ta-W binary alloy system, which evolves from ductile character at Ta-rich compositions to brittleness at high W. We show that a change in slope of the composition-dependent shear modulus near the equiatomic composition coincides with an abrupt change in the Fermi level density of states. We relate the behaviors of the elastic constants to the characters of occupied electronic orbitals close to the Fermi level. Finally, we consider additional alloy systems from groups V and VI and show that qualitatively similar behavior occurs more broadly.
\end{abstract}
\maketitle

\section{Introduction}
Tantalum and tungsten are favored materials for applications that demand high melting temperatures. Tantalum is preferred for applications that demand ductility, while tungsten is favored for its high strength. Naturally, many investigators have examined the mechanical performance of Ta$_{1-x}$W$_x$ alloys~\cite{Wang2020,Duan2022,Chen2023}, and many have also investigated the elastic properties computationally~\cite{Liu2021,Hu2025} in an effort to explain the mechanical behavior.

While no general rigorous principle can predict ductility or brittleness of a metal~\cite{Thompson}, Pugh~\cite{Pugh} noted an empirical relation between the elastic and plastic properties dependent on the ratio of bulk to shear modulus, $P=K/G$. His argument suggested that a low shear modulus would reduce the energetic cost of dislocations, improving malleability, while a high bulk modulus lends strength against fracture. These ideas were further developed by Rice and Thomson~\cite{RiceThomson}. As the Poisson ratio $\sigma$ is a monotonically increasing function of $P$, a high Pugh ratio in turn implies a high Poisson ratio, which has also been taken as a predictor of ductility~\cite{Gao2008}. Structures with cubic symmetry have three independent elastic constants, denoted $\Cone$, $\Ctwo$ and $\C44$, with the bulk modulus $K=(\Cone+2\Ctwo)/3$, and two independent shear moduli, $\C44$ and $\Cprime=(\Cone-\Ctwo)/2$. An isotropic shear modulus $G$ is defined through the Voigt-Reuss-Hill average. Positive Cauchy pressure, $C^{\prime\prime}=\Ctwo-\C44$ has also been considered as an indicator of ductility~\cite{Pettifor,Eberhart,Senkov2021}.

In the free electron approximation~\cite{Kittel} the valence electron density $n$ governs the bulk modulus, with
\begin{equation}
  \label{eq:bulk}
  K = \frac{2}{3} n E_F
\end{equation}
where the Fermi energy
\begin{equation}
  \label{eq:EF}
  E_F = \frac{\hbar^2}{2m}(3\pi^2 n)^{2/3}.
\end{equation}
Shear moduli cannot be predicted from free electron theory, and also transition metal $d$-orbitals are not well described by free electron theory, so accurate prediction of elastic constants requires a higher level of treatment, such as electronic density functional theory~\cite{HohenbergKohn,KohnSham}.

\begin{figure}[b!]
  \centering
  \includegraphics[width=0.45\textwidth,clip]{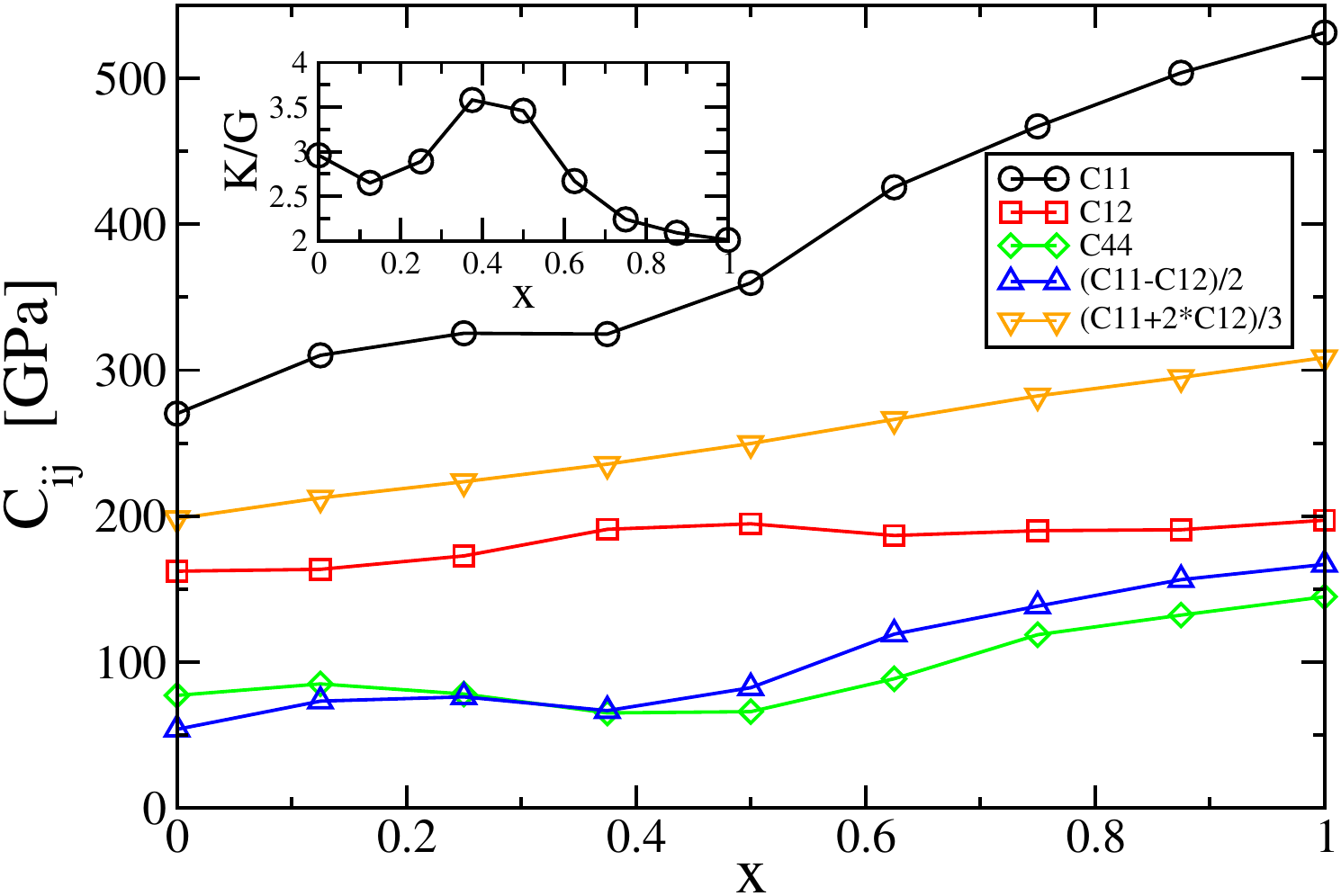}
  \caption{\label{fig:Cij} Cubic elastic constants of Ta-W at various compositions. The Pugh ratio $P=K/G$ is shown in the inset.}
\end{figure}

A recent investigation~\cite{Kareem} of Ta-W alloyed with a low concentration of Nb revealed striking trends in the elastic constants with varying W concentration, $x$. The shear moduli decreased slightly for $x<1/2$ followed by an upturn for $x>1/2$. $\Cone$ increased slowly for $x<1/2$ then more rapidly for $x>1/2$. The bulk modulus, in contrast, enjoyed a steady, nearly linear increase with $x$. Consequently, the Pugh ratio exhibited a broad peak from $x=0$ to $1/2$, followed by a steep decline. This was taken as an indication of ductile behavior for Ta-rich ternary alloys with Nb, and brittleness for W-rich concentrations. Fig~\ref{fig:Cij} presents our elastic constants for the Ta-W binary, which display similar behavior as in the ternary.

This paper lays out the case for an electronic structure-based explanation of the composition-dependence of the elastic constants. We start by considering the electronic densities of states and their underlying band structures. This is done within a rigid band model that allows us to cover the full composition range using only data collected at $x=1/2$. The model is justified in Appendix~\ref{app:rigid} and in Appendix~\ref{app:universal} we address its universality extending to other alloy systems of groups V and VI. Crystal orbital Hamilton population (COHP) analysis~\cite{Dronskowski93} is employed to discriminate between bonding and anti-bonding orbitals. The spatial structure of individual wavefunctions is then analyzed in order to argue for their contributions to different elements of $\Cij$ and to interpret their variation with $x$. The impact of chemical order is briefly addressed. Our findings provide some mechanistic explanation for the observed ``valley of brittleness''~\cite{Poon} along with other observations~\cite{Pettifor,Eberhart,Pettifor1992,Samanta,Asta} that link ductility with electronic structure.

\section{Methods}
Our calculations use the {\tt VASP} code~\cite{KJ_PAW} in the PBE generalised gradient approximation~\cite{Perdew1996} with the recommended Ta$_{\rm pv}$ and W$_{\rm sv}$ PAW potentials at an elevated energy cutoff of 330 eV using ``Accurate'' precision. Structures of different compositions all have 16 atoms, share cubic symmetry, and are described in Appendix~\ref{app:structures}. All structures are fully relaxed at high $k$-point densities. Elastic constants are calculated using a finite difference method. Densities of states are calculated using tetrahedron integration. We use {\tt LOBSTER}~\cite{Lobster} to evaluate the COHPs, and a $\sqrt{2}\times\sqrt{2}$ supercell was employed to allow both nearest- and next nearest-neighbor bonds to be represented. {\tt WaveTrans}~\cite{WaveTrans} is employed to visualize individual wavefunctions. Fermi surface plots were created using {\tt VASPkit}~\cite{vaspkit}.

\section{Density of states and band structure}
\label{sec:dos-bands}

\begin{figure}[b!]
  \centering
\includegraphics[width=0.45\textwidth,clip]{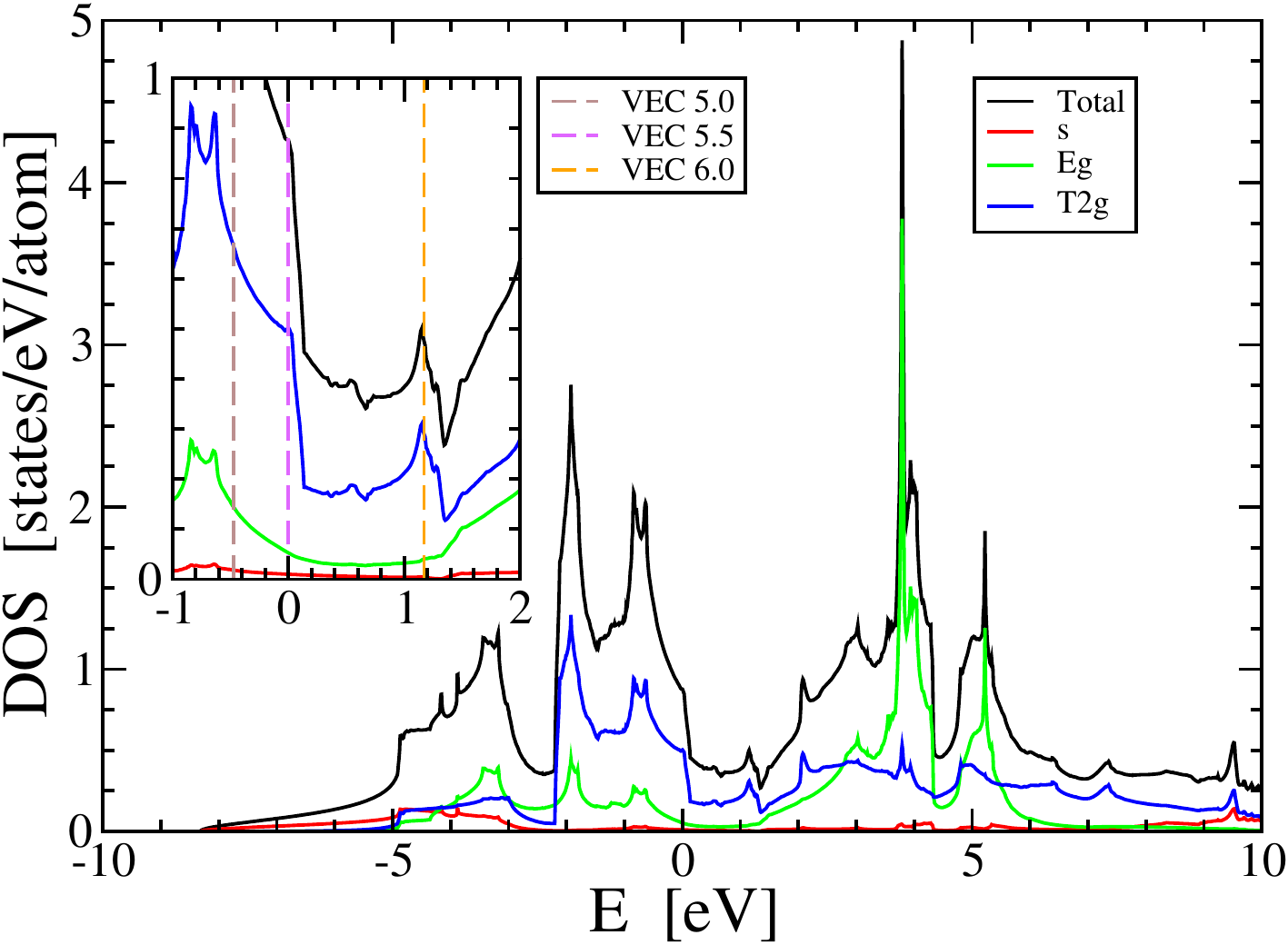}
\caption{\label{fig:dos} (a) Total and partial densities of states of TaW (assuming Pearson type cP2, Strukturbericht B2) with Fermi energy shifted to 0; inset marks Fermi levels corresponding to valence electron counts for elemental Ta, equiatomic TaW, and elemental W.}% (b) Electronic band structure of TaW. Plotting point marker sizes indicate the projections onto atomic orbitals.}
\end{figure}

We begin by inspecting the density of states of TaW in the simple cubic cP2 (B2) structure. Fig.~\ref{fig:dos} displays the full and orbital projected DOS, with the Fermi level shifted to $E=0$. The $d$-orbitals are summed according to irreducible representations of the cubic group, with $\{d_{x^2-y^2}$, $d_{z^2}\}$ contributing to $\Eg$, and $\{d_{xy}, d_{xz}, d_{yz}\}$ contributing to $\T2g$. Note the high shoulder on the DOS below $\EF$, with a deep pseudogap above $\EF$.

The rigid band model assumes that the band structure and DOS are invariant under changes in composition, while the Fermi level shifts according to the valence electron count (VEC). We assign valence 5 to Ta and 6 to W so that TaW has VEC=5.5 per atom. The Fermi levels for elemental Ta and W are marked in Fig.~\ref{fig:dos} relative to the Fermi level of TaW. For low W concentrations, $0\le x \le 1/2$ and $5 \le {\rm VEC} \le 5.5$, the Fermi level DOS is high and is dominated by the $\T2g$ orbitals, while for high W concentrations, $1/2\le x \le 1$ and $5.5 \le {\rm VEC} \le 6$, the Fermi level DOS drops abruptly but again the $\T2g$ orbitals strongly dominate.

\begin{figure}[b!]
  \centering
  \includegraphics[width=0.45\textwidth,clip]{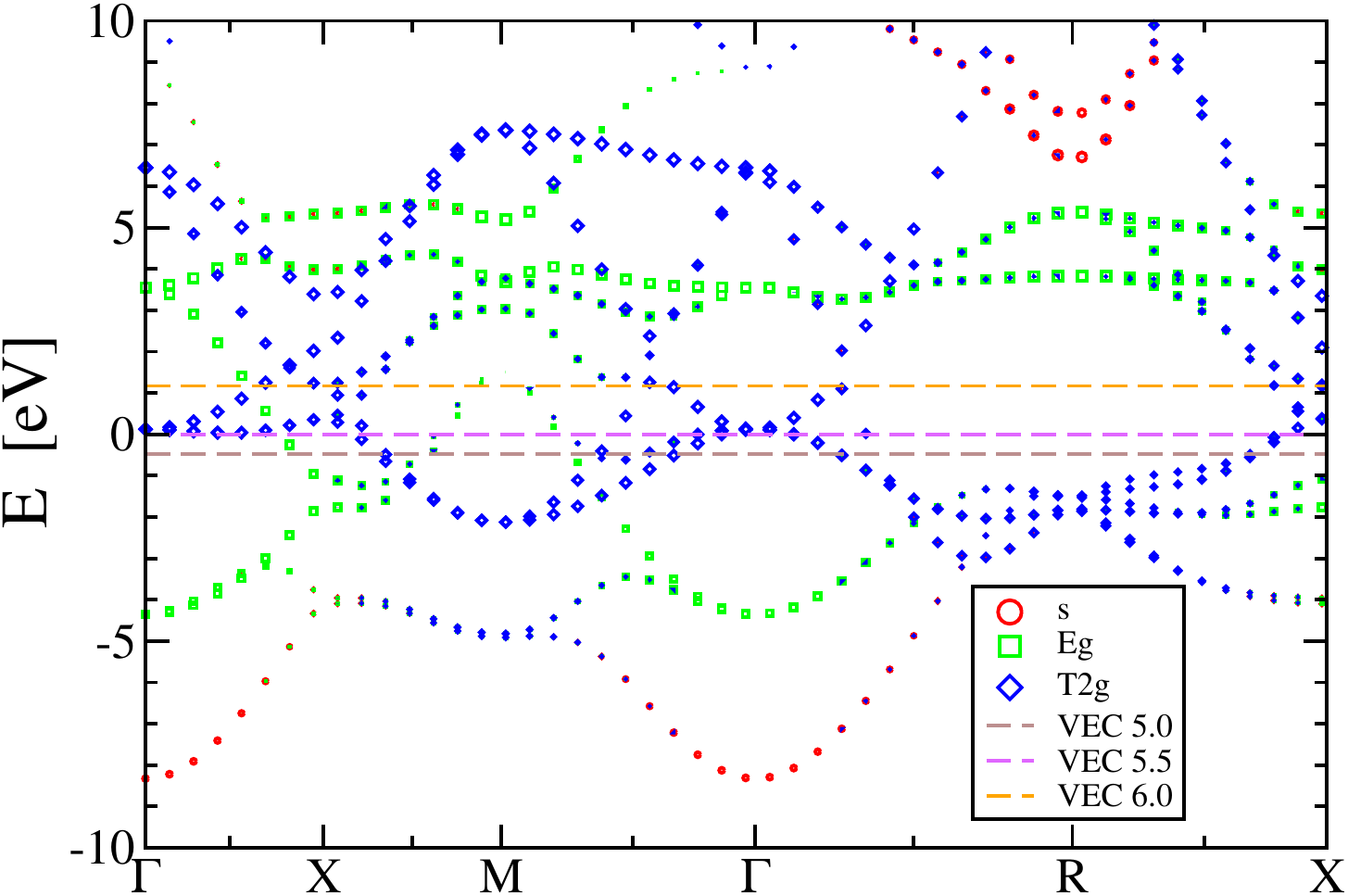}
  \includegraphics[width=0.23\textwidth, trim=14cm 0cm 14cm 0cm, clip]{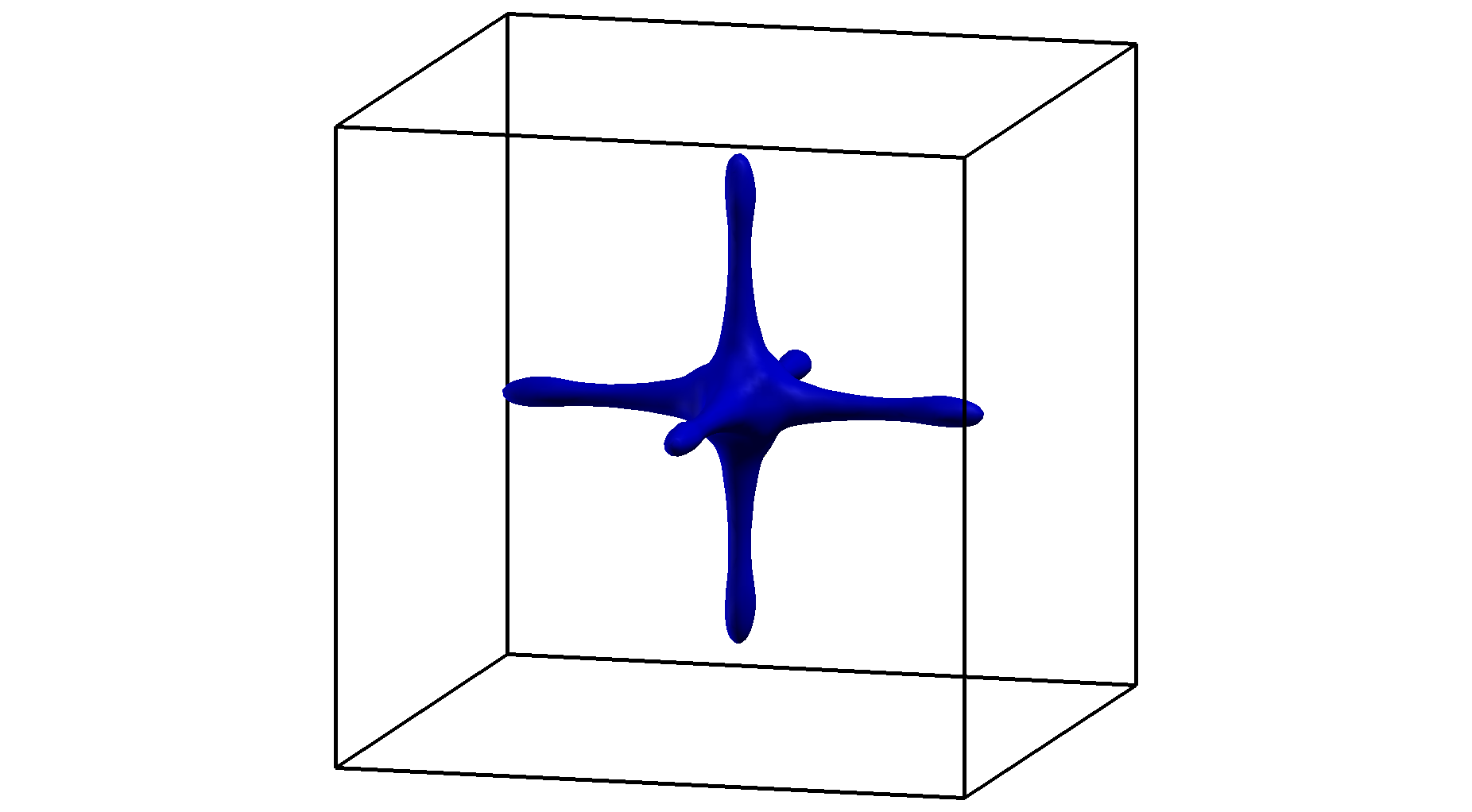}
  \includegraphics[width=0.23\textwidth, trim=14cm 0cm 14cm 0cm, clip]{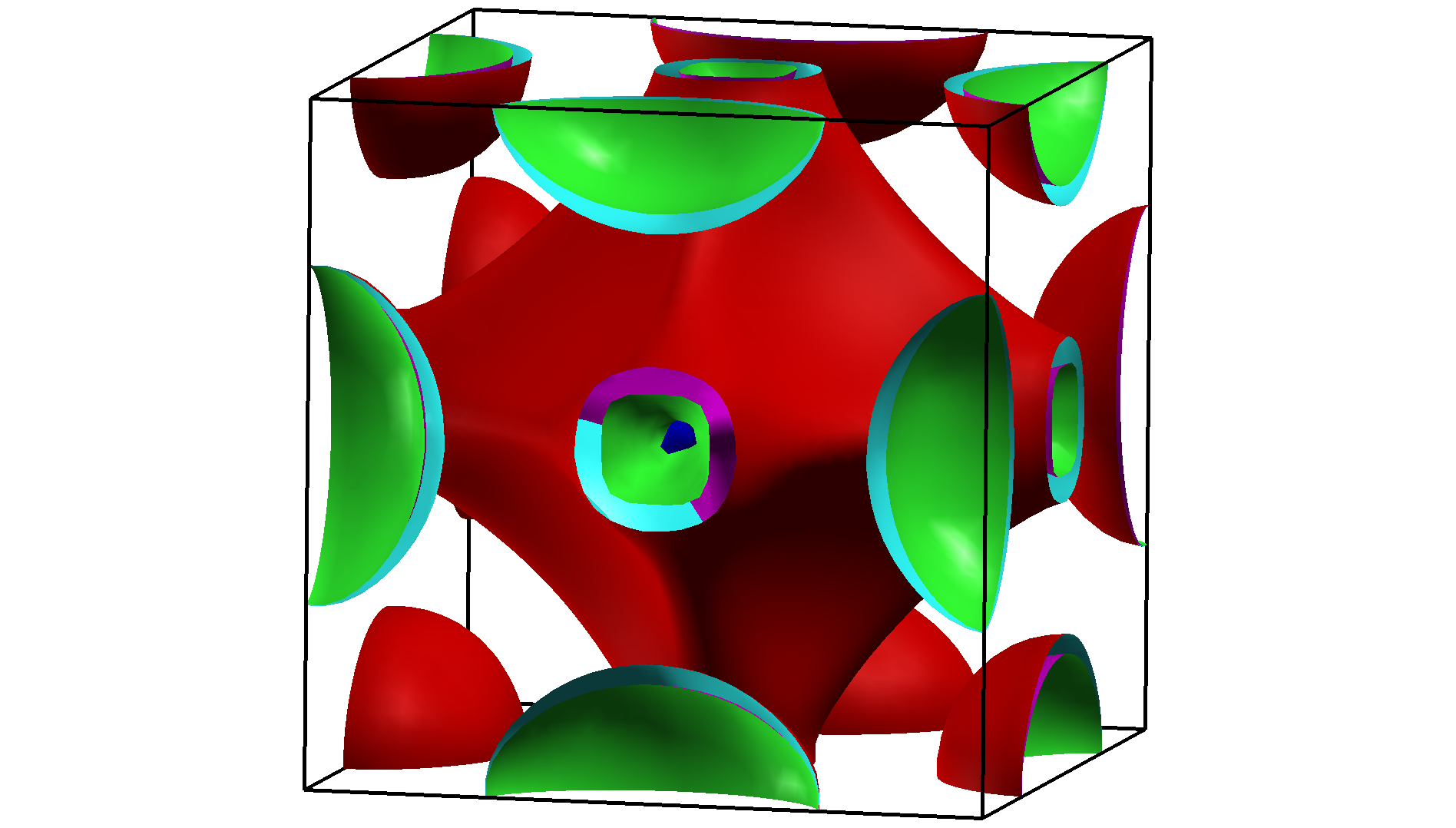}
\caption{\label{fig:bands} (top) Electronic band structure of TaW.cP2. Plotting point marker sizes indicate the projections onto atomic orbitals. (bottom) Fermi surface of flat band (left) and three merged bands (right).}
\end{figure}

Individual bands are plotted within the simple cubic Brillouin zone in Fig.~\ref{fig:bands}. In this ``fat bands'' representation, the sizes of the plotting symbols represent the magnitudes of the projections of each band onto the associated orbitals. The predominance of $\T2g$ orbitals in the VEC interval from 5.0 to 6.0 is clear, although a highly dispersive $\Eg$ band crosses this range along $\Gamma-X$. The band structure explains the sharp dropoff of the DOS near VEC=5.5. All three $\T2g$ bands are dispersive for $E<0$. They meet with three-fold degeneracy and zero slope at the $\Gamma$ $k$-point, as required by symmetry, at an energy just $0.12$ eV above $\EF$. For $E>0$, the $\dyz$ band rises dispersively along $\Gamma-X$ to $1.2$ eV, while $\dxy$ and $\dxz$ remain nearly flat, rising only to $0.35$ eV. Similar behavior is seen along $\Gamma-M$ and $\Gamma-R$.

Three bands contribute to the Fermi surface at VEC 5.5. One surface surrounds the $\Gamma$ point (Brillouin zone center) with narrow extensions along the four-fold axes (\eg $\Gamma-X$). The other two bands are almost perfectly nested, forming channels along $\Gamma-X$ and pockets surrounding the $M$-points.

\section{Crystal orbital Hamilton population}
\label{sec:cohp}

\begin{figure}[t!]
  \centering
  \includegraphics[width=0.45\textwidth,clip]{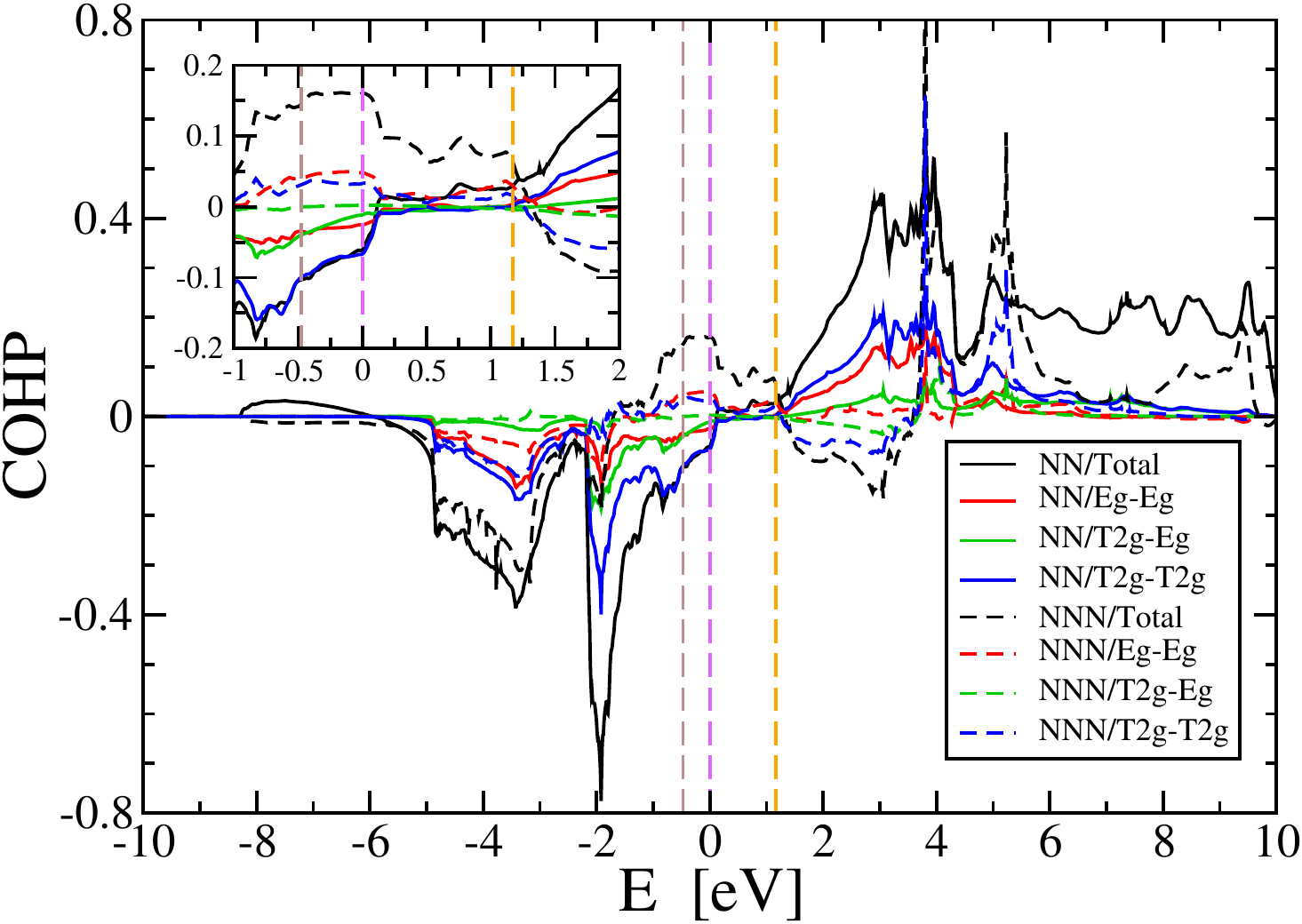}
  \caption{\label{fig:cohp} Selected crystal orbital Hamilton populations between the Ta and W atom of TaW.cP2. Energies corresponding to VEC=5.0, 5.5 and 6.0 are marked as in Fig.~\ref{fig:bands}. Solid curves are for nearest neighbor (NN) bonds, while dashed curves are for next nearest-neighbors (NNN).}
\end{figure}

To obtain deeper insight into the relationship between electronic structure and mechanical properties, we calculate the crystal orbital Hamilton populations (COHPs~\cite{Dronskowski93,Lobster}). The COHP is effectively a measure of covalent interatomic bonding strength weighted by the joint partial densities of states. Negative values are bonding and positive values are anti-bonding. Integrated COHPs can be taken as heuristic measures of covalent bonding strength. As is evident in Fig.~\ref{fig:cohp}, the total COHP for the nearest neighbor bond is strongly bonding for low VEC $x<1/2$ and weakly {\em anti}-bonding for high VEC at $x>1/2$. $\T2g-\T2g$ interactions transition from strongly bonding at low VEC to nearly neutral at high VEC. $\Eg-\Eg$ and $\T2g-\Eg$ interactions transition from weakly bonding to neutral. All of these transitions take place near $E=0.12$ eV, where the $\Gamma$-point degeneracy lies.

At the next nearest-neighbors, the bonding character is nearly reversed. The total NNN bond strength transitions from strongly {\em anti}-bonding at low VEC to weakly anti-bonding. Both $\T2g-\T2g$ and $\Eg-\Eg$ interactions contribute to the NNN anti-bonding character.

\section{Wavefunctions}
\label{sec:wavefunctions}

\begin{figure}[h!]
  \centering
  \includegraphics[width=0.45\textwidth, trim=0.7cm 1cm 2cm 2cm, clip]{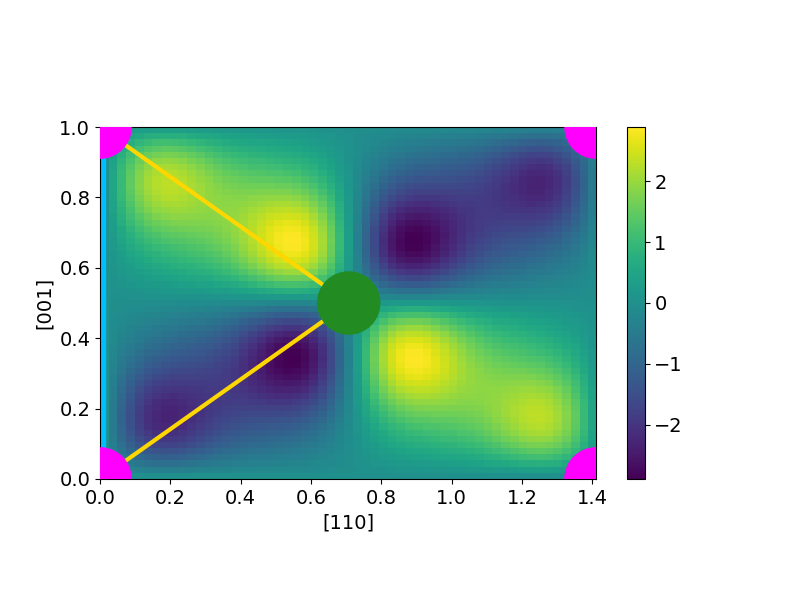}
  \includegraphics[width=0.45\textwidth, trim=6cm 0cm 4cm 0cm, clip]{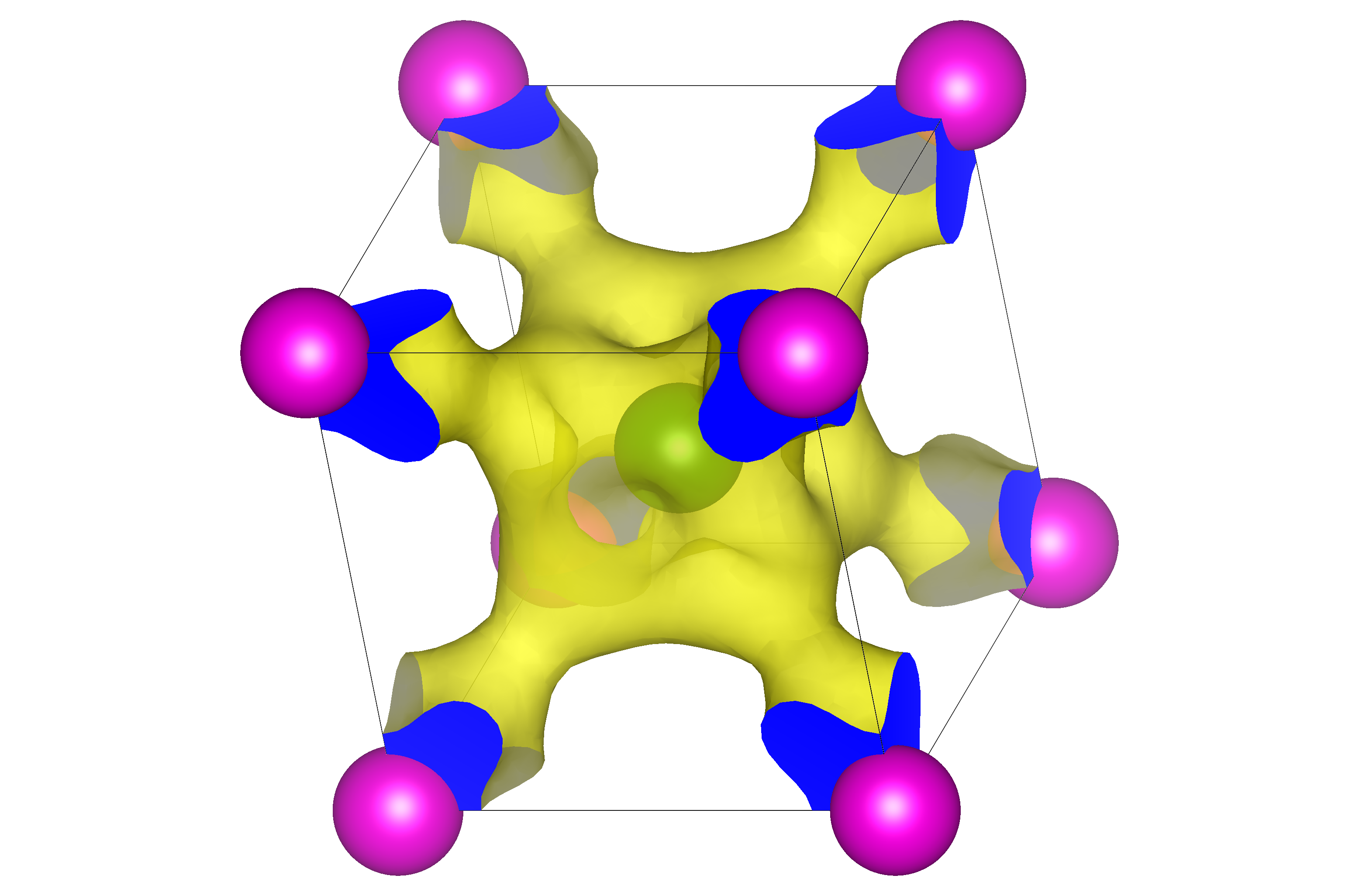}
   \includegraphics[width=0.45\textwidth, trim=4cm 3cm 4cm 3cm, clip]{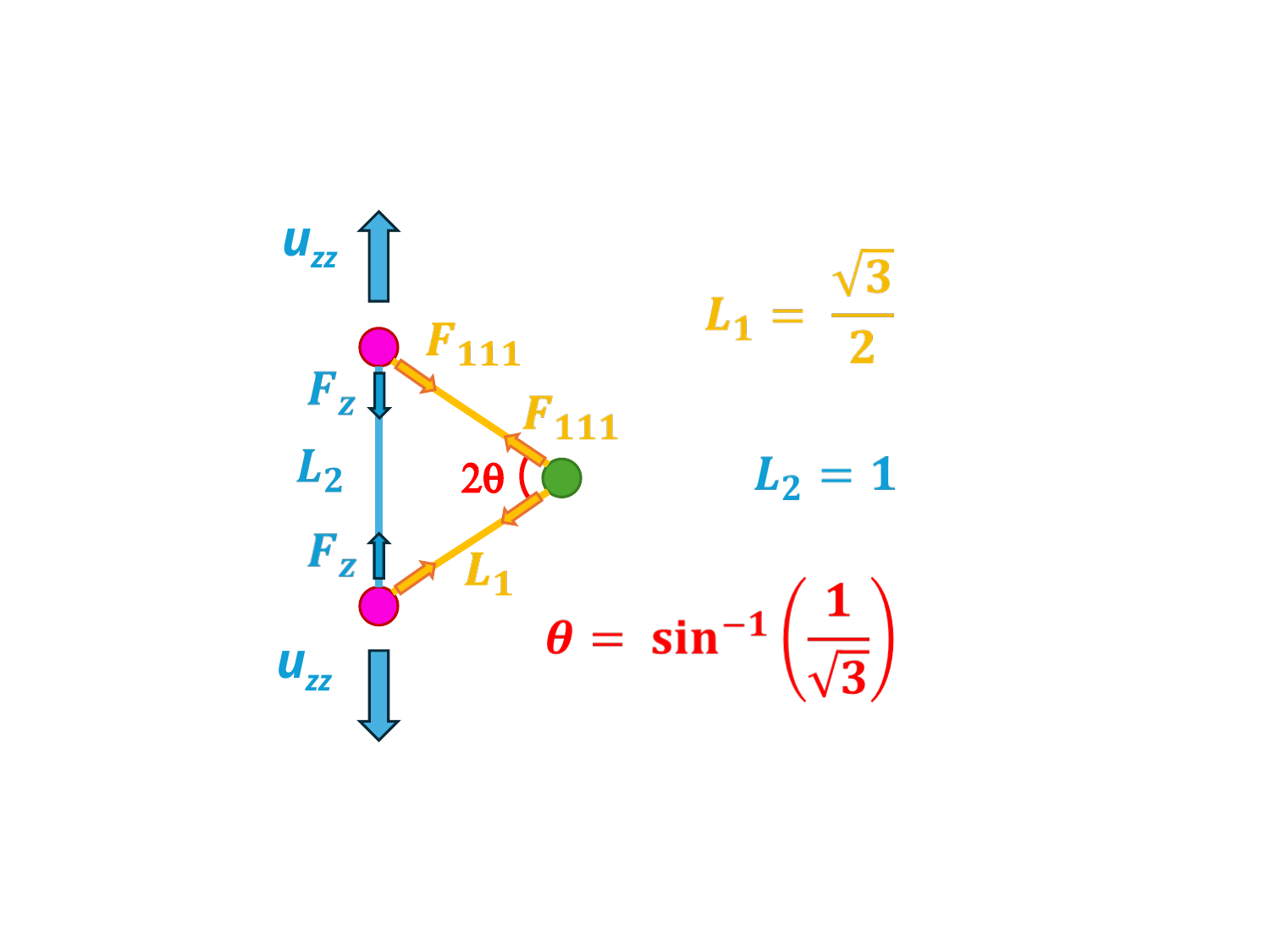}
  \caption{\label{fig:wave} (top): $\Gamma$-point $\dxz$ orbital of TaW.cP2 at energy $E=0.12$ eV. The horizontal [110] axis and vertical [001] axis are labeled. The Ta atom position is marked in magenta and the W atom in turquoise. The nearest neighbor interatomic bond lying along the [111] direction is marked in gold, while the next nearest-neighbor lying along the [001] axis at $x=0$ is marked in blue. (middle): Partial charge in range $E\in(-0.474,0.12)$. The isosurface is at $5\times 10^{-3}$ electrons per~\AA$^{3}$. (bottom): Bonds and their associated forces. }
\end{figure}

The geometrical properties of individual wavefunctions can provide mechanistic insight into the energetics of deformation~\cite{Bojun,Vishnu}, and hence into the elasticity. These considerations also help to interpret the COHPs. Our analysis is aided by the importance of states at the $\Gamma$ $k$-point, which are purely real-valued, making them easy to analyze. These are the states whose occupancies change near VEC=5.5 and $E=0.12$ eV, where the elastic anomalies occur. Two of the three $\T2g$ orbitals are occupied at $x=1/2$. These orbitals are bonding in character, as is evident from inspection of the wavefunction in the (110) plane (see Fig.~\ref{fig:wave}). Covalent bonding occurs when the signs of the wavefunction match along the length of the bond. The covalent character is further illustrated by the partial charge density of all states that lie in the energy range $E\in(-0.474,0.12)$ that extends from VEC 5.0 to 5.5.

Consider uniaxial strain $u_{zz}$ applied along the [001] axis. This stretches the nearest-neighbor bonds that lie in the $\langle 111\rangle$ directions, and also increases the bond angle $2\theta$; both effects create stress of type $\sigma_{zz}$ (force along [001]) contributing to $\Cone$, and they also create stresss of types $\sigma_{xx}$ and $\sigma_{yy}$ (forces along [100] and [010]) contributing to $\Ctwo$. Now consider the next-nearest-neighbor bonds that lie in the $\langle 001 \rangle$ directions and contribute exclusively to $\Cone$. The wavefunction illustrated in Fig.~\ref{fig:wave} (top) vanishes along the [001] bond, and in general it reverses sign with increasing $z$, making it {\em anti}-bonding, and thus inhibiting the growth of $\Cone$ as VEC increases. Once the VEC passes 5.5, two of the three $\T2g$ orbitals have become filled and the anti-bonding COHP strength is reduced, so $\Cone$ grows more rapidly. At the same time, the flattening of $\Ctwo$ with increasing VEC beyond 5.5 may be caused in part by the limited growth of the $\T2g-\T2g$ bond occupation together with the vanishing COHP strength. The opposing trends of $\Cone$ and $\Ctwo$ allow the bulk modulus $K=(\Cone+2\Ctwo)/3$ grows steadily with increasing $x$, as expected from the steady increase of electron density $n$ (see Eqs.~(\ref{eq:bulk}) and~(\ref{eq:EF})). In contrast, the tetragonal shear modulus $\Cprime=(\Cone-\Ctwo)/2$ is nearly flat for $x<1/2$ and then grows rapidly for $x\gtrsim 1/2$.

Shear strain of type $u_{xy}$, which create stress $\sigma_{xy}$ = $\C44 u_{xy}$, compresses half of the nearest-neighbor bonds while stretching the other half, resulting in weak contributions to $\C44$. It only impacts the bond angles and the next-nearest neighbor bond lengths at second order, yielding no additional contribution to the (linear) elastic moduli. This may explain why $\C44$ approximately tracks the composition-dependence of $\Cprime$, but with a lower total variation.

\section{Chemical order}
\label{sec:order}

\begin{figure}[h!]
  \centering
  \includegraphics[width=0.45\textwidth,clip]{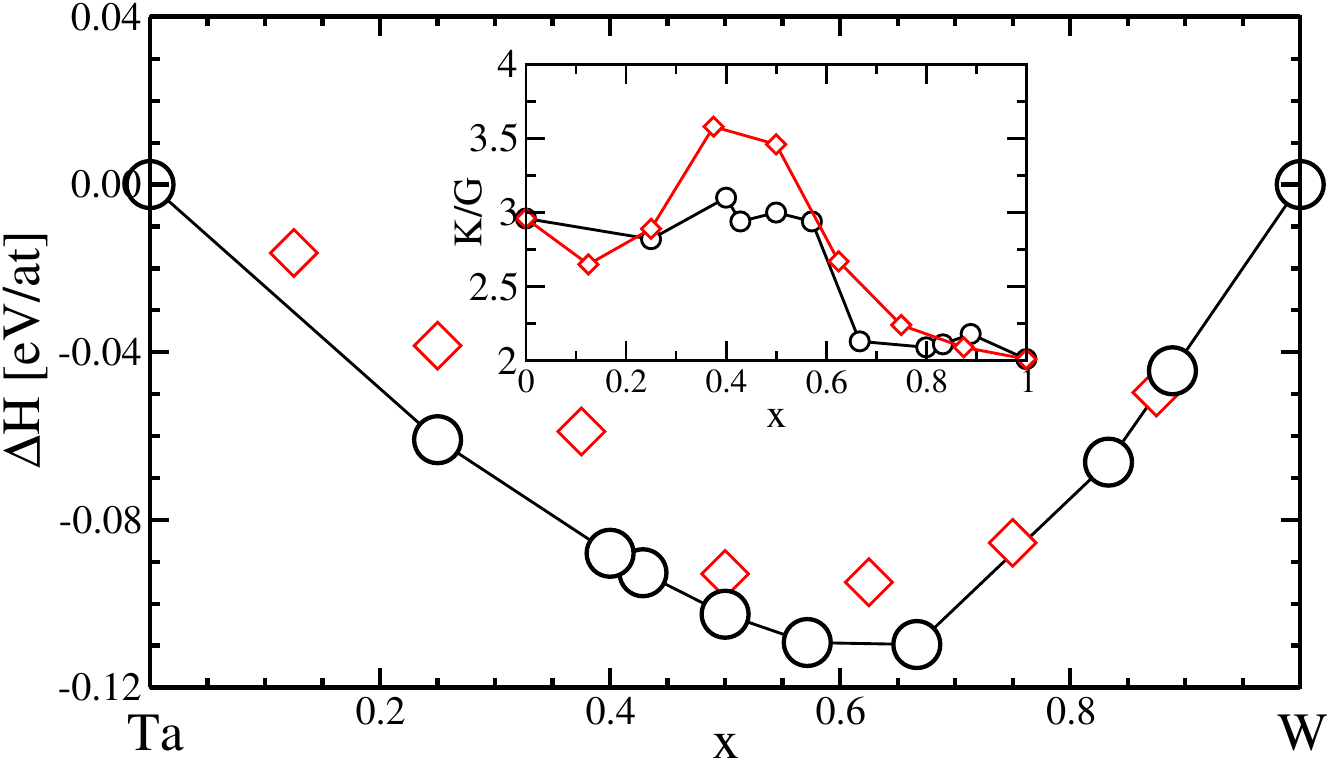}
  \caption{\label{fig:hull} Formation enthalpies of ground state (black) and cubic (red) structures. Ground state structures are: Ta3W (Pearson type mC16); Ta3W2 (hR5); Ta4W3 (hR7); TaW (oA12); Ta3W4 (hR7); TaW2 (tI6); TaW4 (tI10); TaW5 (mC12); TaW8 (tI18). Pugh ratios $P=K/G$ are shown in the inset for ground state (black) and cubic (red) structures.}
\end{figure}

The Ta-W structures considered above were designed to preserve cubic symmetry without regard to their energies. To observe the impact of chemical order on the elasticity, energy minimizing arrangements of Ta and W on BCC lattice sites were obtained~\cite{JOM} using the {\tt ATAT} toolkit~\cite{ATAT,VANDEWALLE2009266}. Fig.~\ref{fig:hull} displays the formation enthalpies of the ground state structures that lie on the convex hull of enthalpy {\em vs.} composition. For $0<x<1/2$ our cubic structures lie above the hull, while for $x\geq 1/2$ they lie closer to the hull. Because the ground state structures are non-cubic, we do not show the individual $\Cij$, but the Pugh ratios of Voigt-Reuss-Hill orientation-averaged  moduli are given in the inset. In general, the ground state structure Pugh ratios fall below those of the cubic structures. This is likely due to elevated shear moduli associated with enhanced frequencies of Ta-W bonds in the ground state structures. The effect on the Pugh ratio is strongest for $0<x<1/2$, similar to the behavior of the enthalpies.

\section{Conclusion}
\label{sec:conclude}

In conclusion, we suggest that the anomalous composition-dependence of the Ta$_{1-x}$W$_x$ elastic constants originates in the electronic structure. Our analysis invoked the structure of the density of states, electronic band structure, crystal orbital Hamiltonian population, and electronic wave functions. From VEC=5.0 at $x=0$ up to VEC=5.5 at $x=1/2$, electrons enter covalent bonds between nearest neighbor atoms formed by $\T2g$ orbitals. The increased bond strength leads to rising values of both $\Cone$ and $\Ctwo$. However, the increase of $\Cone$ is tempered by the $\T2g$ {\em anti}-bonds between next nearest neighbors. Beyond $x\gtrsim 1/2$, both the nearest-neighbor bonding and the next nearest-neighbor anti-bonding are diminished, as shown by reduced COHP magnitudes, causing a flattening in the composition-dependent trend of $\Ctwo$ and a corresponding increased slope of $\Cone$. $\C44$ follows a trend similar to $\Cprime$.  A consequence of these trends is that the Pugh ratio $P=K/G$ exhibits a broad peak for $x\lesssim 1/2$. The Cauchy pressure $C^{\prime\prime}=\Ctwo-\C44$ is positive for all $x\in(0,1)$, but peaks over the same range of $x$ as the Pugh ratio. Both of these indicators suggest that Ta-rich alloys with W may be more ductile than W-rich alloys.

We invoked a rigid band model to explain the behavior of Ta-W, and we showed that the model is justified across varying concentrations within the Ta-W alloy system. The spirit of the rigid band model suggests that similar effects could occur for other alloys within the families of group V and VI elements (see Appendix~\ref{app:universal}). Indeed, we find qualitatively similar behavior with Pugh ratios peaking at $0<x<1/2$, but with substantial quantitative differences among peak positions and heights. Likewise the rigid band model holds less well for the densities of states between the different alloy systems than it does within a single system of variable concentration.

\begin{acknowledgments}
  KA and JK were supported by the Naval Nuclear Laboratory under award No. 1047622 for the elasticity investigation of the Nb-Ta-W ternary and its relevance to ductility. MW was supported by the Department of Energy under grant No. DE-SC0014506 for the interpretation of electronic structure-related elasticity.
\end{acknowledgments}

\begin{appendix}
  \renewcommand\thefigure{\thesection.\arabic{figure}}
  \renewcommand\thetable{\thesection.\roman{table}}

\section{Justification of rigid band model}
\label{app:rigid}
\setcounter{figure}{0}

\begin{figure}[h!]
  \centering
  \includegraphics[width=0.45\textwidth, clip]{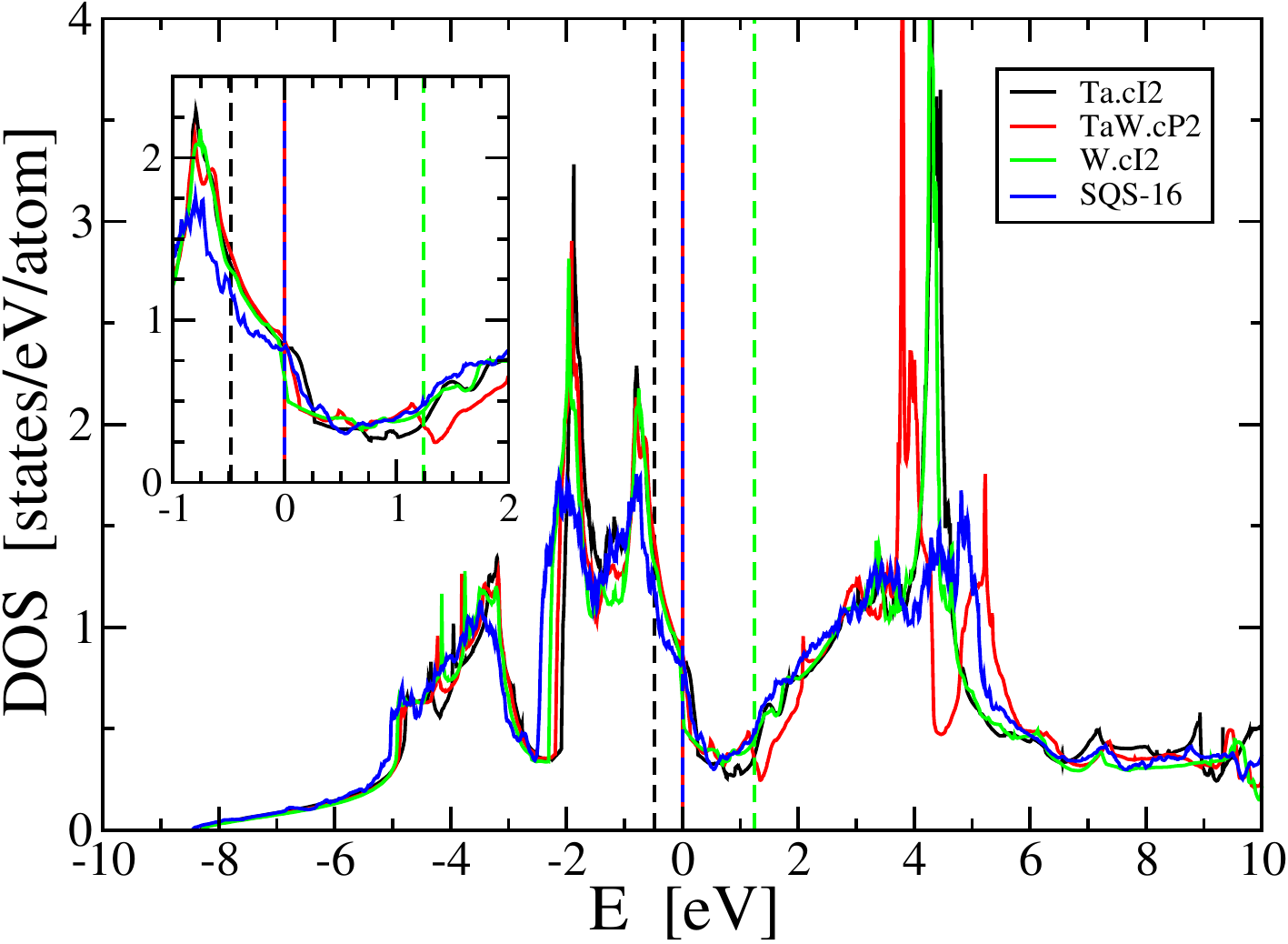}
  \caption{\label{fig:rigid} Densities of states of Ta.cI2, W.cI2, TaW.cP2, and a 16-atom equiatomic TaW SQS. Energies have been shifted to place VEC=5.5 at 0; dashed lines indicate the actual Fermi energies.}
\end{figure}

To verify the applicability of the rigid band model, we compare the calculated densities of states across several compositions and structures including a special quasirandom structure (SQS) \cite{sqs-zunger}. The rigid band model needs to pass two tests: Are the bands and densities of states sufficiently robust against changes in composition? Does the chemically ordered cP2 structure accurately represent the disordered cI2 solid solution? Fig.~\ref{fig:rigid} answers these questions affirmatively for Ta-W. All calculations utilized the tetrahedron integration method. Energies have been shifted to align the densities of states so that VEC=5.5 per atom (obtained from the integrated DOS) sits at $E=0$. A Gaussian smearing of width 0.1 eV has been applied to the SQS for enhanced clarity. Overall, the DOS resemble each other over the entire range of both occupied and unoccupied states, independently of composition and structure. The shoulder from VEC 5.0-5.5 and the pseudogap from VEC 5.5-6.0 match especially closely.

\section{Universality}
\label{app:universal}
\setcounter{figure}{0}

\begin{figure}[h!]
  \centering
  \includegraphics[width=0.45\textwidth, clip]{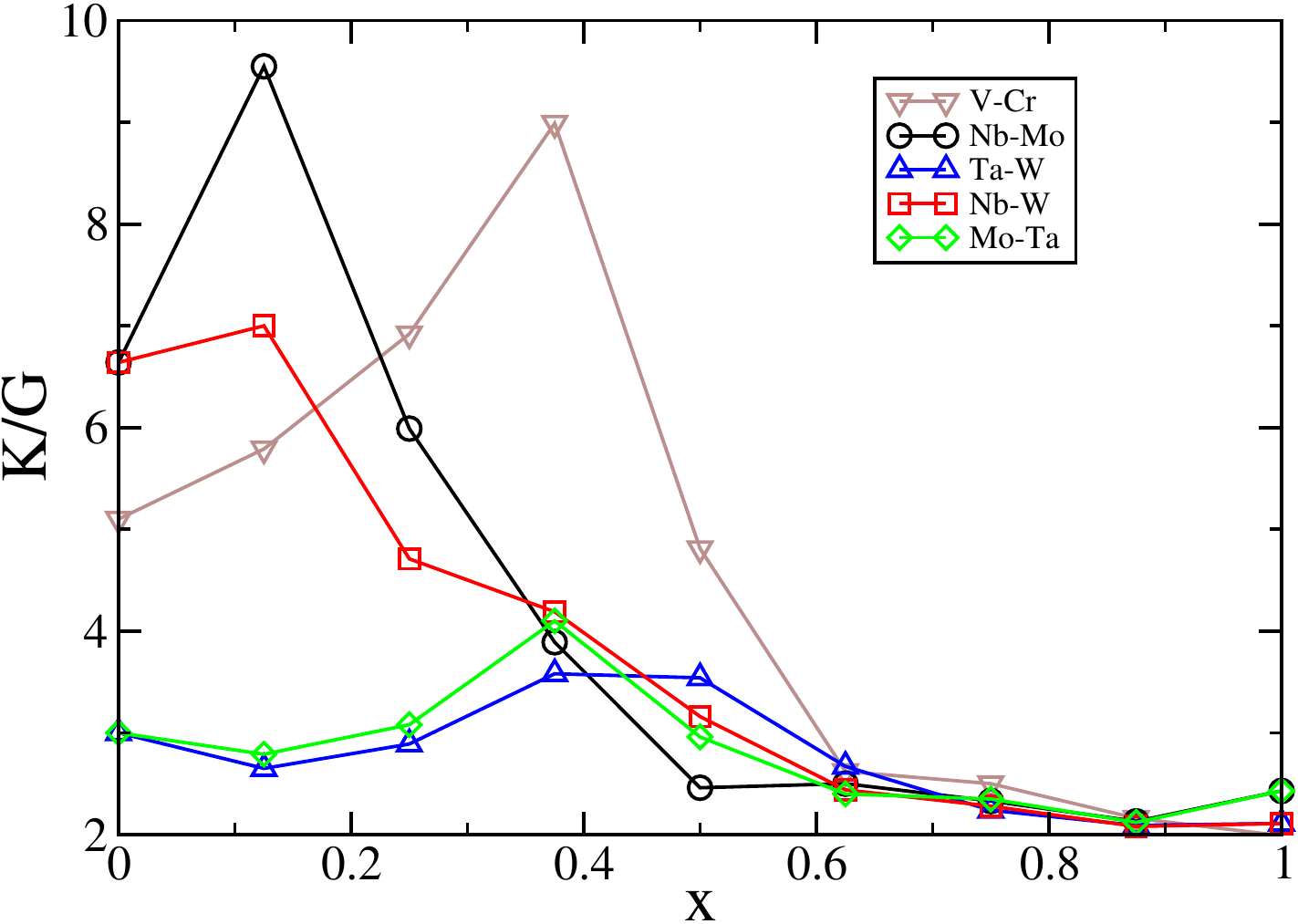}
  \caption{\label{fig:Pugh} Composition-dependent Pugh ratios of selected group V-VI solid solutions.}
\end{figure}

The rigid band model suggests that similar electronic structures and trends in elasticity should occur universally for a variety of other equiatomic group V-VI compounds. To test this, we have calculated the composition-dependent Pugh ratios of selected binaries (Fig.~\ref{fig:Pugh}). VCr, NbMo, and TaW represent the first, second, and third rows of transition metals, respectively; NbW and TaMo cross between the second and third rows. In all cases, the Pugh ratios peak at concentrations below $1/2$ and fall off at higher $x$. However, there are quantitative differences among the alloy systems.
\begin{figure}[h!]
  \centering
  \includegraphics[width=0.45\textwidth, clip]{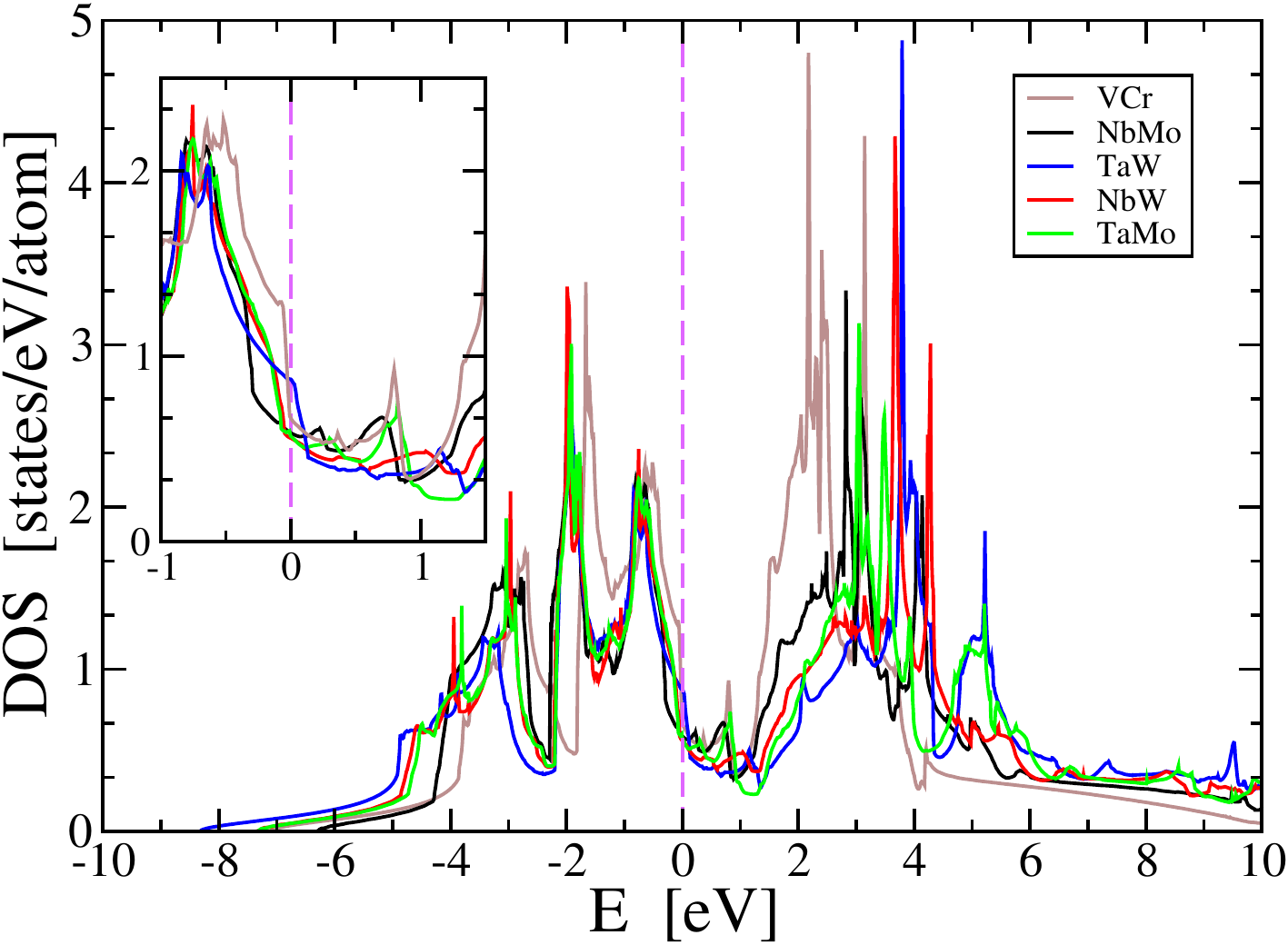}
  \caption{\label{fig:dos-V-VI} DOS of selected group V-VI binaries in the cP2 structure. All DOS shifted to place $\EF$ at $E=0$.}
\end{figure}

The peak concentrations seem to correlate with the onsets of pseudogaps in the corresponding DOS that we present in Fig~\ref{fig:dos-V-VI}.  The main DOS features of interest remain, with a high shoulder on the left of VEC=5.5, and a sharp rise to the right of VEC=6.0. Some deviations are evident in the precise shapes of the shoulders and pseudogaps. On the whole, the rigid band model works less well across different alloy systems than it does among Ta-W alloys of varying compositions.

\section{Structures considered}
\label{app:structures}
\setcounter{figure}{0}

Elasticity calculations for Ta$_{1-x}$W$_x$ were performed for the following structures. We give the chemical formulas, relaxed lattice constants, Pearson types, strukturbericht where available, space group number, and Wyckoff sites.

\subsection{Cubic structures}
\begin{itemize}
\item $x=0$: Ta$_2$, $a=3.318$\AA, Pearson type cI2, strukturbericht A2, group 229 (Ta at site 2a)
\item $x=1/8$: Ta$_{14}$W$_2$, $a=6.586$\AA, Pearson type cI16, group 229 (Ta at sites 6b and 8c; W at site 2a)
\item $x=1/4$: Ta$_3$W, $a=6.553$\AA, Pearson type cF16, strukturbericht D0$_3$, group 225 (Ta at sites 4b and 8c; W at site 4a)
\item $x=3/8$: Ta$_{10}$W$_6$, $a=6.510$\AA, Pearson type cI16, group 229 (Ta at sites 2a and 8c; W at 6b)
\item $x=1/2$: TaW, $a=3.241$\AA, Pearson type cI2, strukturbericht B2, group 221 (Ta at site 1a; W at site 1b)
\item $x=5/8$: Ta$_6$W$_{10}$, $a=6.449$\AA, Pearson type cI16, group 229 (Ta at site 6b; W at sites 2a and 8c)
\item $x=3/4$: TaW$_3$, $a=6.420$\AA, Pearson type cF16, strukturbericht D0$_3$, group 225 (Ta at site 4a; W at sites 4b and 8c)
\item $x=7/8$: Ta$_2$W$_{14}$, $a=6.392$\AA, Pearson type cI16, group 229 (Ta at site 2a; W at sites 6b and 8c)
\item $x=1$: W$_2$, $a=3.183$\AA, Pearson type cI2, strukturbericht A2, group 229 (W at site 2a)
\end{itemize}

%For the $\Eg$ wavefunction calculation we employed a $\sqrt{2}\times\sqrt{2}$ supercell rotated by 45$^\circ$ with structure Ta$_2$W$_2$, $a=4.70$, $b=4.72$, and $c=3.33$\AA, Pearson type oC4, group 65 (Ta at site 2a, W at site 2c).

\subsection{Ground state structures}
\begin{itemize}
    \item $x=0.25$: Ta$_3$W, $a=10.43, b=3.27, c=9.27$~\AA, $\beta=116.7^\circ$, Pearson type mC16, group 12 (Ta1@4i (0.62,0,0.44); Ta2@4i (0.62,0,0.93); Ta3@4i (0.13,0,0.19); W@4i (0.13,0,0.69))
    \item $x=0.4$: Ta$_3$W$_2$, $a=b=4.6, c=14.2$~\AA, $\gamma=120^\circ$, Pearson type hR15, group 166 (Ta1@6c (0,0,0.21); Ta2@3a (0,0,0); W@6c (0,0,0.40))
    \item $x=0.43$: Ta$_4$W$_3$, $a=4.6, c=19.7$~\AA, $\gamma=120^\circ$, Pearson type hR7, group 160 (Ta1@3a (0,0,0,001), Ta2@3a (0,0,0,0.43), Ta3@3a (0,0,0,0.57), Ta4@3a (0,0,0,0.85), W1@3a (0,0,0,0.15), W2@3a (0,0,0,0.28), W3@3a (0,0,0,0.71))
    \item $x=0.5$: TaW, $a=13.8, b=3.2, c=4.6$~\AA, Pearson type oC12, group 65 (Ta1@4g (0.66,0,0); Ta2@2c (1/2,0,1/2); W1@4h (0.17,0,1/2); W2@2a (0,0,0))
    \item $x=0.57$: Ta$_3$W$_4$, Pearson type hR7, group 166 (Ta1@3a (0,0,0); Ta2@6c (0,0,0.43); W1@6c (0,0,0.15); W2@6c (0,0,0.28))
    \item $x=0.67$, TaW$_2$, $a=3.2, c=9.7$~\AA, Pearson type tI6, group 139 (Ta@2a (0,0,0), W@4e (0,0,0.67))
    \item $x=0.8$, TaW$_4$, $a=3.2, c=16.1$~\AA, Pearson type tI10, group 139 (Ta@2a (0,0,0); W1@4e (0,0,0.60); W2@4e (0,0,0.80))
    \item $x=0.83$: TaW$_5$, $a=1.01, b=3.2, c=6.4$~\AA, $\beta=108.5^\circ$, Pearson type mC12, group 12 (Ta@2b (0,1/2,0); W1@4i (0.17,0,0.83); W2@4i (0.17,0,0.33); W3@2d (0, 1/2, 1/2))
    \item $x=0.89$: TaW$_8$, $a=9.6, c=3.2$, Pearson type tI18, group 139 (Ta@2b (0,0,1/2); W1@8j (0.17, 0.5, 0); W2@8h (0.17, 0.17, 0))
\end{itemize}

\end{appendix}

\bibliography{refs}

\end{document}